\documentclass[aps,prl,superscriptaddress]{revtex4}%
\usepackage{amsfonts}
\usepackage{amsmath}
\usepackage{amssymb}
\usepackage{graphicx}
\usepackage{color}
\usepackage{amsmath}
\usepackage{hyperref}
\usepackage{float}
\usepackage{mathtools}
\usepackage{algorithm}
\usepackage{algpseudocode}

\newcommand{\ii}{\textrm{i}}
\newcommand{\Laplace}{\Delta}
\newcommand{\etas}{\vec{\boldsymbol{\eta}}}
\newcommand{\Mat}[1]{\big[{#1}\big]}

\newcommand{\bjacobi}{\texttt{bjacobi}}
\newcommand{\apfc}{\texttt{apfc}}

\def\equationautorefname~#1\null{%
  Eq.~(#1)\null
}
\def\figureautorefname~#1\null{%
  Fig.~#1\null
}
\def\algorithmautorefname~#1\null{%
  Alg.~#1\null
}

\begin{document}
\title{An efficient numerical framework for the amplitude expansion of the phase-field crystal model}
\date{\today}

\author{Simon Praetorius} 
\affiliation{Institute  of Scientific Computing,  Technische  Universit\"at  Dresden,  01062  Dresden,  Germany}
\author{Marco Salvalaglio} \email{marco.salvalaglio@tu-dresden.de} 
\affiliation{Institute  of Scientific Computing,  Technische  Universit\"at  Dresden,  01062  Dresden,  Germany}
\author{Axel Voigt}
\affiliation{Institute  of Scientific Computing,  Technische  Universit\"at  Dresden,  01062  Dresden,  Germany}
\affiliation{Dresden Center for Computational Materials Science (DCMS), TU Dresden, 01062 Dresden, Germany}

\begin{abstract}
The study of polycrystalline materials requires theoretical and computational techniques enabling multiscale investigations. The amplitude expansion of the phase-field crystal model (APFC) allows for describing crystal lattice properties on diffusive timescales by focusing on continuous fields varying on length scales larger than the atomic spacing. Thus, it allows for the simulation of large systems still retaining details of the crystal lattice. Fostered by the applications of this approach, we present here an efficient numerical framework to solve its equations. In particular, we consider a real space approach exploiting the finite element method. An optimized preconditioner is developed in order to improve the convergence of the linear solver. Moreover, a mesh adaptivity criterion based on the local rotation of the polycrystal is used. This results in an unprecedented capability of simulating large, three-dimensional systems including the dynamical description of the microstructures in polycrystalline materials together with their dislocation networks.
\end{abstract}

\maketitle

\section{Introduction}
The study of polycrystalline materials requires modeling over various length scales. While the nucleation and structure of defects has to be considered on an atomistic resolution, the size, shapes and arrangements of grains can be considered on mesoscopic length scales and mechanical behavior can be tackled on a macroscopic scale. However, all these phenomena strongly affect material properties, such as fracture or yield stress and thus require multiscale investigations \cite{Rollett2015}.

Coarse-grained, mesoscale approaches are therefore viable tools to provide bridging-scale information of polycrystalline materials. Among the many different models reported in the literature, the so-called phase-field crystal (PFC) model has been developed in order to filter out atom vibrations and enabling the investigation of relatively long dynamics \cite{Elder2002,Elder2004,Emmerich2012}. It describes the crystal lattice by means of a continuous field that is the atomic probability density, $n$, averaged over vibrational length scales. Although the original model has been further extended, as e.g. in Refs \cite{Greenwood2010,Huang2010,Kocher2015}, and applied to describe several different properties and mechanisms for crystals and quasicrystals, as e.g. in Refs. \cite{Rottler2012,Berry2014,Achim2014,Hirvonen2016,Wang2016,Yamanaka2017}, it is restricted to relatively small sizes as the continuous density has still to be resolved at the atomic length scale.

This limit has been overcome by the so-called amplitude expansion of the PFC model (APFC), providing coarse-graining in both time and space in a single framework \cite{Goldenfeld2005,Athreya2006,GoldenfeldJSP2006,Yeon2010}. This model focuses on the amplitudes of the atomic probability density which vary on a larger length scale than the atomic spacing. In its standard formulation, it is restricted to relatively small deformations \cite{Spatschek2010} and approximate atomic rearrangements such as dislocation-core structures. However, it encodes a detailed description of deformations matching continuum-elasticity theories and allows for large simulations approaching the macroscopic length scales, still retaining details of the atomic length scale \cite{Salvalaglio2018a}. The original model can be further extended as it has been done, for instance, to account for binary systems \cite{Elder2010} and to control the energy of defects and interfaces \cite{SalvalaglioAPFC2017}. Moreover, it has been recently used to study the anisotropic shrinkage in 3D of small-angle spherical grain boundaries (GBs) regardless of the crystal lattice symmetry \cite{Salvalaglio2018}. In addition, the APFC framework has been proven suitable to allow for the description of hydrodynamics \cite{Heinonen2016}, dislocation dynamics \cite{Skaugen2018} and surface-energy anisotropy \cite{Ofori-Opoku2018} within the more general PFC framework.

In this work, we report on the development of an efficient numerical framework to integrate the partial differential equations of the APFC model, enabling unprecendented simulations of large polycrystalline systems in two- and three-dimensions. We provide an optimized discretization of the APFC equations based on the one reported in Ref.~\cite{SalvalaglioAPFC2017}. Then, we propose a new preconditioner allowing for faster convergence of iterative linear solvers. We also illustrate a mesh-adaptivity strategy exploiting continuous fields, namely local crystal rotations, which can be derived directly by the complex amplitudes to solve for in the APFC model.  Performance studies showing comparisons with a standard solver and assessment of the numerical parameter entering the method are reported. Then, we use the developed framework to simulate the large scale growth of a polycrystal in 2D and 3D accounting for both the evolution of the microstructures and the defects forming between grains having different orientations.

% -----------------------------------------------------------------------------------------------------------
\section{Model and Implementation}
The APFC model \cite{Goldenfeld2005,Athreya2006,GoldenfeldJSP2006,Yeon2010} is based on the representation of the atomic probability density, $n$, as sum of plane waves
\begin{equation}\label{eq:density}
  n(\mathbf{r}) = n_0+\sum_{j=1}^{J} \left[ \eta_j(\mathbf{r}) e^{\ii\mathbf{k}_j \cdot
\mathbf{r}}+ \eta_j^*(\mathbf{r}) e^{-\ii\mathbf{k}_j \cdot \mathbf{r}}\right],
\end{equation}
with $n_0$ the average density, here set to zero without loss of generality, $\eta_j(\mathbf{r})$ the amplitude of each plane wave, and $\mathbf{k}_j$ the reciprocal lattice vector representing a specific crystal symmetry.
Complex amplitude functions $\eta_j:\Omega\rightarrow\mathbb{C}$ with $\Omega$ a rectangular (2D) or parallelepiped (3D) domain are considered, allowing for describing distortions and rotations of the crystal lattice with respect to a reference state described by the $\mathbf{k}_j$ vectors. The model is based on the definition of a free energy in terms of the $\eta_j$'s in the approximation of slowly varying amplitudes, i.e., varying over a length scale significantly larger than the lattice spacing,
\begin{equation}\label{eq:energyamplitude}
  F(\etas) := \int_{\Omega} f^\textrm{s}(\etas,\etas^*) + \frac{1}{2}\sum_{j=1}^J |\mathcal{G}_j \eta_j|^2\,\text{d}\mathbf{r},
\end{equation}
with $\etas=\{\eta_j\}$, $\mathcal{G}_j \coloneqq \Laplace+2\ii\mathbf{k}_j \cdot \nabla$, and $f^\textrm{s}$ a polynomial bulk free energy density. For the sake of convenience, the energy is here normalized by a positive, non-zero parameter $a_0$ which would multiply the second term at the right-hand side of \eqref{eq:energyamplitude} \cite{Elder2010a,SalvalaglioAPFC2017}. The evolution laws for amplitudes $\eta_j$ are given by the $L^2$-gradient flow of the energy $F$,
\begin{equation}\label{eq:amplitudetime}
  \frac{\partial \eta_j}{\partial t} =-\kappa_j \frac{\delta F}{\delta \eta_j^*}\,,\quad j=1,\ldots,J,
\end{equation}
with $\kappa_j=a_0|\mathbf{k}_j|^2$ a mobility coefficient and 
\begin{equation}\label{eq:gs}
  \frac{\delta F}{\delta \eta_j^*} = \mathcal{G}^2_j\eta_j + \frac{\delta f^\textrm{s}(\etas,\etas^*)}{\delta \eta_j^*}.
\end{equation}

The bulk energy term $f^\textrm{s}(\etas,\etas^*)$ entering Eq.~\eqref{eq:energyamplitude} can be written as 
\begin{equation}
  f^\textrm{s}(\etas,\etas^*)=\frac{a_1}{a_0}A+\frac{a_2}{a_0}\left(A^2-2\sum_{j=1}^J |\eta_j|^4\right) + g^\textrm{s}(\etas,\etas^*),
\end{equation}
with $a_i$ positive parameters, $g^\textrm{s}(\etas,\etas^*)$ a polynomial in $\etas$ and $\etas^*$ encoding the lattice symmetry, and $A=2\sum_{j=1}^J |\eta_j|^2$. In this work we consider a triangular lattice in 2D and a face-centered cubic (FCC) lattice in 3D. The former is obtained by setting $J=3$, $g^{\rm s}$ entering Eq.~\eqref{eq:gs} as
\begin{equation}
  g^\textrm{s}(\etas,\etas^*)=-2\frac{a_3}{a_0}\left(\eta_1\eta_2\eta_3 + \eta_1^*\eta_2^*\eta_3^*\right)\,,
\end{equation}
and the set of $\mathbf{k}_j$ vectors as
\begin{equation}
  \mathbf{k}_1=\left( -\frac{\sqrt{3}}{2},-\frac{1}{2}\right), \qquad  \mathbf{k}_2=\left(0,1\right.), \qquad  \mathbf{k}_3=\left(\frac{\sqrt{3}}{2},-\frac{1}{2}\right).
\end{equation}
For the FCC lattice $J=7$, thus leading to a larger number of amplitude functions to be considered. For the sake of brevity we refer to Ref.~\cite{SalvalaglioAPFC2017} for the corresponding $g^{\rm s}$ polynomial and $\mathbf{k}_j$ vectors. Following Refs.~\cite{Elder2010a,SalvalaglioAPFC2017,Salvalaglio2018}, the parameters entering the energy are set to favor the growth of the solid phase as $a_0=0.98$, $a_1=0.01$, $a_2=0.25$, $a_3=0.5$.

% --------------------------------------------------
\subsection{Discretization and Nonlinear solver}
Throughout the following, we use the $L^2(\Omega,\mathbb{C})$ scalar product $(u,\,v) \coloneqq \int_\Omega u^*(\mathbf{r})\cdot v(\mathbf{r})\,\text{d}\mathbf{r}$, with $u,v$ complex scalar or vector valued functions and $u^*$ the complex conjugate of $u$. By introducing a set of auxiliary variables $\zeta_j \coloneqq \mathcal{G}_j\eta_j$, we derive the weak form of the evolution equations \eqref{eq:amplitudetime} using a natural splitting of the operator $\mathcal{G}_j^2$:
\begin{equation}\label{eq:weak_evolution}\begin{split} 
  (\zeta_j,\, \vartheta_1) + (\nabla\eta_j,\,\nabla\vartheta_1) - 2\ii(\mathbf{k}_j\cdot\nabla\eta_j,\,\vartheta_1) &= 0 \\
  \left(\frac{\partial \eta_j}{\partial t},\,\vartheta_2\right) - \kappa_j\big((\nabla\zeta_j,\,\nabla\vartheta_2) - 2\ii(\mathbf{k}_j\cdot\nabla\zeta_j,\,\vartheta_2)\big) &= \left(\frac{\delta f^\textrm{s}(\etas,\etas^*)}{\delta \eta_j^*},\,\vartheta_2\right)\,,\qquad\forall\vartheta_1,\vartheta_2\in H^1(\Omega,\mathbb{C})
\end{split}\end{equation}
This set of coupled second order equations can then be discretized using conforming finite-dimensional approximations of the functions space $H^1(\Omega,\mathbb{C})$. Therefore, let us now consider a triangulation of the domain $\Omega$, $\mathcal{T}_h$, with $h$ the minimal diameter of the grid elements, and the finite-element space $\mathcal{V}_h\subset H^1(\Omega,\mathbb{C})$,
\[\begin{split}
  \mathcal{V}_h &= \{\vartheta\in C(\Omega,\mathbb{C})\,:\, \vartheta|_S\in\mathbb{P}_1(S,\mathbb{C}),\,S\in\mathcal{T}_h\}\,,
\end{split}\]
the continuous space of complex valued local linear polynomials on elements $S\in\mathcal{T}_h$ of the triangulation. The space discretization method of \eqref{eq:weak_evolution} corresponds to find $\eta^h_j,\zeta^h_j\in L^1([0,T], \mathcal{V}_h)$ such that
\begin{equation}\label{eq:discrete_evolution}\begin{split} 
  (\zeta^h_j,\, \vartheta^h_1) + (\nabla\eta^h_j,\,\nabla\vartheta^h_1) - 2\ii(\mathbf{k}_j\cdot\nabla\eta^h_j,\,\vartheta^h_1) &= 0 \\
  \left(\frac{\partial \eta^h_j}{\partial t},\,\vartheta^h_2\right) - \kappa_j\big((\nabla\zeta^h_j,\,\nabla\vartheta^h_2) - 2\ii(\mathbf{k}_j\cdot\nabla\zeta^h_j,\,\vartheta^h_2)\big) &= \left(\frac{\delta f^\textrm{s}(\etas^h,\etas^{h*})}{\delta \eta_j^*},\,\vartheta^h_2\right)\,,\qquad\forall\vartheta^h_1,\vartheta^h_2\in \mathcal{V}_h\,,
\end{split}\end{equation}
for $j=1,\ldots,J$ in the time interval $[0,T]$ provided that $\eta_j^h(t=0)=\eta_j^{h,0}$ is given. For better readability, we drop the superscript $h$ in the following. 

% -----------------------------------------------------------------------------------------------------------
Due to the bulk polynomial $f^\textrm{s}$, equations \eqref{eq:discrete_evolution} are a set of nonlinear equations. We combine a simplified Newton method with the idea of an operator splitting method, in order to separate the equation for each amplitude. This will allow us to efficiently solve the set of equations also if the number of amplitudes is large, as in the case of FCC symmetry. We introduce the Jacobian of $F^\textrm{s}_j \coloneqq \delta f^\textrm{s}/\delta\eta_j^*$, but only in direction $\eta_j$, i.e.,
\[
  dF^\textrm{s}_j(\etas) \coloneqq \frac{\delta F^\textrm{s}_j}{\delta\eta_j}(\etas) = \frac{\delta^2 f^s}{\delta\eta_j\delta\eta_j^*}(\etas)\,.
\]

Let $\{\phi_i\}$ be a basis of $\mathcal{V}_h$. Moreover, let us introduce the mass matrix $\mathbf{M} \coloneqq \Mat{(\phi_i,\,\phi_j)}$, the stiffness matrix $\mathbf{K} \coloneqq \Mat{(\nabla\phi_i,\nabla\phi_j)}$, $\mathbf{T}^k \coloneqq \Mat{(2\mathbf{k}_k\cdot\nabla\phi_i,\phi_j)}$, $\mathbf{J}^k(\etas) \coloneqq \Mat{(-dF^\textrm{s}_k(\etas)\phi_i,\,\phi_j)}$, and the vector $\mathbf{h}^k(\etas) \coloneqq \Mat{(F^\textrm{s}_k(\etas) - dF^\textrm{s}_k(\etas)\eta_k,\,\phi_j)}$. Note, that these matrices are assembled with the real valued basis functions $\phi_i$. 

For the time discretization we choose a backward Euler method. In the following let $0=t_0 < t_1 < \ldots < t_M = T$ be a discretization of the time interval $[0,T]$ with $\tau_m :=t_{m}-t_{m-1}$ and the timestep solution $\eta^m_j = \sum_i \eta^m_{j,i}\phi_i \equiv \eta_j(t_m)$ with $\boldsymbol{\eta}_j^m=\Mat{\eta^m_{j,i}}$ the vector of the coefficients for the $j$-th amplitude at the timestep $m$. Analogously, we define $\zeta^m_j = \sum_i \zeta^m_{j,i}\phi_i \equiv \zeta_j(t_m)$ and $\boldsymbol{\zeta}_j^m=\Mat{\zeta^m_{j,i}}$. 

\begin{algorithm}[ht]
\begin{algorithmic}
\State Let $\boldsymbol{\eta}^0_j,\boldsymbol{\zeta}^0_j\in\mathbb{C}^{{\rm dim}(V_h)}\;(j=1,\ldots,J)$ be given. \Comment{Initial solution}
\For{$m=1,\ldots,M$} \Comment Loop over all timesteps
  \State Let $\boldsymbol{\eta}^{(0)}_j=\boldsymbol{\eta}^{m-1}_j$ and $\boldsymbol{\zeta}^{(0)}_j=\boldsymbol{\zeta}^{m-1}_j\;(j=1,\ldots,J)$. \Comment{Initial iteration for Newton method}
  \For {$k=1,\ldots,K$} \Comment{Newton iteration}
    \For {$j=1,\ldots,J$} \Comment{Loop over amplitudes}
        \State Find $\boldsymbol{\eta}^{(k)}_j\in\mathbb{C}^{{\rm dim}(V_h)},\,\boldsymbol{\zeta}^{(k)}_j\in\mathbb{C}^{{\rm dim}(V_h)}$, such that
\begin{equation}\label{eq:linear_system1}
  \begin{bmatrix}
  \mathbf{M} & \mathbf{K} - \ii\mathbf{T}^j \\
  -\kappa_j(\mathbf{K} - \ii\mathbf{T}^j) & \frac{1}{\tau_m}\mathbf{M} + \mathbf{J}^j(\etas^{(k-1)})
  \end{bmatrix}\begin{pmatrix}
  \boldsymbol{\zeta}^{(k)}_j \\ \boldsymbol{\eta}^{(k)}_j
  \end{pmatrix} = \begin{pmatrix}
  0 \\ \frac{1}{\tau_m}\mathbf{M}\boldsymbol{\eta}^{m-1}_j + \mathbf{h}^j(\etas^{(k-1)})
  \end{pmatrix}\,.
\end{equation}
    \EndFor
  \EndFor
  \State Update $\boldsymbol{\eta}^{m}_j=\boldsymbol{\eta}^{(K)}_j$ and $\boldsymbol{\zeta}^{m}_j=\boldsymbol{\zeta}^{(K)}_j\; (j=1,\ldots,J)$.
\EndFor
\end{algorithmic}
\caption{\label{alg:nonlinear_solver}Timestepping scheme and nonlinear solver for the APFC equation}
\end{algorithm}

The timestep iteration with an inner Picard iterative process for the linearized $F^\textrm{s}_j$ can be found in \autoref{alg:nonlinear_solver}.
The discretization is implemented in the finite-element framework AMDiS \cite{Vey2007, Witkowski2015}.

% -----------------------------------------------------------------------------------------------------------
\subsection{Solving the Linear System}
The skew symmetric linear system \eqref{eq:linear_system1} needs to be solved in each timestep and in each Newton iteration. For large-scale numerical setups in 3D, the widely used direct spare LU factorizations are out of applicability due to large memory requirements. Therefore, we concentrate on preconditioned iterative methods, belonging in particular to the Krylov subspace, to solve \eqref{eq:linear_system1}. Two strategies will be compared: domain decomposition with local sparse LU factorization also known as block Jacobi (\bjacobi) preconditioner \cite{Saad2003} and a dedicated preconditioner based on the schur-complement method (\apfc), as detailed in the following.

For the construction of the schur-complement preconditioner, we follow the lines of \cite{Boyanova2012,Axelsson2013}, and first simplify the notation by neglecting the amplitude index $j$. For each amplitude a structurally similar system has to be solved, so we can construct one preconditioner that can be applied to all amplitudes, with adapted coefficients. Also, we drop the timestep index $m$ and Newton iterate index $(k)$ and just write $\tau\equiv\tau_m$. In spite of the operator $\mathcal{G}_j$ we introduce $\mathbf{G}^j \coloneqq -\mathbf{K} + \ii\mathbf{T}^j$ and write the linear system as
\begin{equation}\label{eq:linear_system2}
  \underbrace{\begin{bmatrix}
  \mathbf{M} & -\mathbf{G} \\
  \kappa\mathbf{G} & \frac{1}{\tau}\mathbf{M} + \mathbf{J}
  \end{bmatrix}}_{\mathbf{A}}\underbrace{\begin{pmatrix}
  \boldsymbol{\zeta} \\ \boldsymbol{\eta}
  \end{pmatrix}}_{\mathbf{x}} = \begin{pmatrix}
  0 \\ \frac{1}{\tau}\mathbf{M}\boldsymbol{\eta}^{m} + \mathbf{h}
  \end{pmatrix}
\end{equation}
with schur-complement $\mathbf{S} = \frac{1}{\tau}\mathbf{M} + \mathbf{J} + \kappa\mathbf{G}\mathbf{M}^{-1}\mathbf{G}$.

While $\mathbf{M}$ and $\mathbf{K}$ are real matrices, $\mathbf{G}$, $\mathbf{J}$, and $\mathbf{h}$ may be complex valued. Following the ideas of \cite{Boyanova2012,Axelsson2013} to approximate the schur-complement matrix $\mathbf{S}$ in a way that allows for a simple factorization, we propose the preconditioner:
\begin{equation}\label{eq:preconditioner}
  \mathbf{P} := \begin{bmatrix}
  \mathbf{M} & 0 \\
  \kappa\mathbf{G} & \frac{1}{\sqrt{\tau}}\mathbf{M} + \sqrt{\kappa}\mathbf{K}
  \end{bmatrix}\begin{bmatrix}
  \mathbf{I} & -\mathbf{M}^{-1}\mathbf{K} \\
  0 & \mathbf{M}^{-1}(\frac{1}{\sqrt{\tau}}\mathbf{M} + \sqrt{\kappa}\mathbf{K})
  \end{bmatrix} = \begin{bmatrix}
  \mathbf{M} & -\mathbf{K} \\
  \kappa\mathbf{G} & \frac{1}{\tau}\mathbf{M} + 2\sqrt{\frac{\kappa}{\tau}}\mathbf{K} + \kappa(\mathbf{K}\mathbf{M}^{-1}\mathbf{K} - \mathbf{G}\mathbf{M}^{-1}\mathbf{K})
  \end{bmatrix}
\end{equation}
with schur complement $\mathbf{S}_P \coloneqq (\frac{1}{\sqrt{\tau}}\mathbf{M} + \sqrt{\kappa}\mathbf{K})\mathbf{M}^{-1}(\frac{1}{\sqrt{\tau}}\mathbf{M} + \sqrt{\kappa}\mathbf{K})$. Since the highest order term in $\mathbf{G}\mathbf{M}^{-1}\mathbf{G}$ corresponds to $\mathbf{K}\mathbf{M}^{-1}\mathbf{K}$, the matrix $\mathbf{S}_P$ is expected to be a good approximation of $\mathbf{S}$. 

The application of $\mathbf{P}$ to a vector $\mathbf{b}=(\boldsymbol{\beta}_1,\boldsymbol{\beta}_2)^\top\in\mathbb{C}^{2\cdot\operatorname{dim}(V_h)}$, i.e., $\mathbf{x}=\mathbf{P}_2^{-1}\mathbf{b}$, with $\mathbf{x}=(\boldsymbol{\chi}_1,\boldsymbol{\chi}_2)^\top$, can be performed in four steps:
\begin{enumerate}
\item solve $\mathbf{M} \boldsymbol{\beta}'_1 = \boldsymbol{\beta}_1$
\item solve $(\frac{1}{\sqrt{\tau}}\mathbf{M} + \sqrt{\kappa}\mathbf{K})\boldsymbol{\beta}'_2 = \boldsymbol{\beta}_2 - \kappa \mathbf{G}\boldsymbol{\beta}'_1$
\item solve $(\frac{1}{\sqrt{\tau}}\mathbf{M} + \sqrt{\kappa}\mathbf{K})\boldsymbol{\chi}_2 = \mathbf{M}\boldsymbol{\beta}'_2$
\item $\boldsymbol{\chi}_1 = \boldsymbol{\beta}'_1 - \frac{1}{\sqrt{\kappa}}(\boldsymbol{\beta}'_2 - \frac{1}{\sqrt{\tau}}\boldsymbol{\chi}_2)$.
\end{enumerate}
Thus, only one mass-matrix and two diffusion-matrix systems need to be solved, plus two matrix-vector products need to be performed in the application of the preconditioner. Since efficient (iterative) linear solvers exist to approximate the solution of $\mathbf{M}^{-1}\mathbf{b}$ and $(\frac{1}{\sqrt{\tau}}\mathbf{M} + \sqrt{\kappa}\mathbf{K})^{-1}\mathbf{b}$, the preconditioner is cheap to apply.

\begin{figure}[ht] % Fig. 1
\begin{center}
\includegraphics[scale=1]{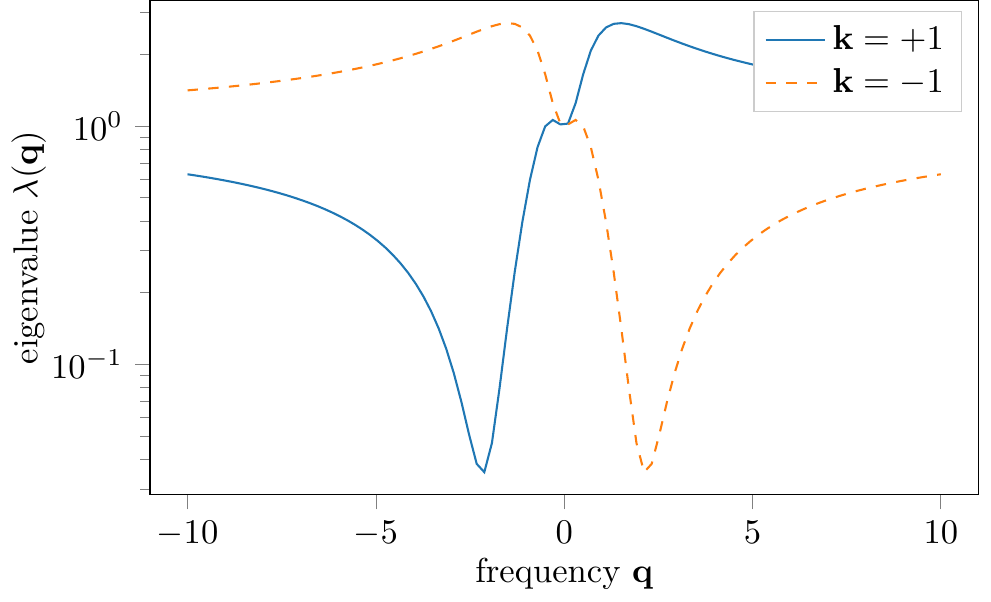} 
\caption{\label{fig:figure1}Visualization of the operator spectrum for $\mathbf{k}\in\{-1,+1\}$.}
\end{center}
\end{figure}

In order to analyze the quality of the preconditioner $\mathbf{P}$, we look at the eigenvalue spectrum of the right-preconditioned linear system $\mathbf{A}\mathbf{P}^{-1}$. Instead of evaluating the eigenvalues directly, we consider the symbols of the differential operators and the resulting spectrum for a simplified system. Therefore, let $\Omega$ be a periodic one-dimensional domain of length $2\pi$. We set the nonlinear term to $\mathbf{J}=0$, and consider only lattice vectors $\mathbf{k}\in\{-1,+1\}$. Let $\mathcal{F}$ denote the non-unitary Fourier transform in angular frequency $\mathbf{q}$, i.e.,
\[
  \widehat{\eta}(\mathbf{q}) \coloneqq \mathcal{F}\eta = \int_\Omega \eta(\mathbf{r})e^{-i\mathbf{q}\cdot\mathbf{r}}\,\text{d}\mathbf{r}\,,
\]
with inverse transform $\mathcal{F}^{-1}$ and the linear operators $\mathbf{A}$ and $\mathbf{P}$ expressed as $\mathbf{A}\mathbf{x} = \mathcal{F}^{-1}(\mathcal{A}\widehat{\mathbf{x}})$ and $\mathbf{P}\mathbf{x} = \mathcal{F}^{-1}(\mathcal{P}\widehat{\mathbf{x}})$, respectively, with symbols $\mathcal{A}$ and $\mathcal{P}$. 
In terms of the frequency $\mathbf{q}$, these symbols can be written as
\[
  \mathcal{A} = \begin{bmatrix}
  1 & \mathbf{q}^2 + 2\mathbf{q}\cdot\mathbf{k} \\
  -\kappa(\mathbf{q}^2 + 2\mathbf{q}\cdot\mathbf{k}) & \frac{1}{\tau}
  \end{bmatrix},\qquad\mathcal{P} = \begin{bmatrix}
  1 & 0 \\
  -\kappa(\mathbf{q}^2 + 2\mathbf{q}\cdot\mathbf{k}) & \frac{1}{\sqrt{\tau}} + \sqrt{\kappa}\mathbf{q}^2
  \end{bmatrix}\begin{bmatrix}
  1 & -\mathbf{q}^2 \\
  0 & \frac{1}{\sqrt{\tau}} + \sqrt{\kappa}\mathbf{q}^2
  \end{bmatrix}\,.
\]
For $\mathcal{Q}:=\mathcal{A}\mathcal{P}^{-1}$, $\kappa=1$, and $\tau=1$, we obtain the spectrum
\[
  \sigma(\mathcal{Q}) = \left\{1, \lambda(\mathbf{q})\coloneqq\frac{\mathbf{q}^4 \mp 4\mathbf{q}^3 + 4\mathbf{q}^2 + 1}{\mathbf{q}^4 + 2\mathbf{q}^2 + 1} \;\big|\;\mathbf{q}\in\mathbb{R} \right\}\,,
\]
where the sign of the second term in the nominator depends on the choice of the k-vector, see \autoref{fig:figure1} for a visualization.
The eigenfunction $\lambda$ approaches the value 1 for zero and $\pm \infty$ frequencies and has a minimum $\min\{\lambda(\mathbf{q})\,|\,\mathbf{q}\in\mathbb{R}\} \approx 0.035$ and maximum $\max\{\lambda(\mathbf{q})\,|\,\mathbf{q}\in\mathbb{R}\} \approx 2.70$. The operator symbol $\mathcal{A}$ forms a normal matrix for $|\mathbf{k}_j|=1$ and the preconditioned operator spectrum is in the positive and real interval $[0.035, 2.7]$, leading to an asymptotic convergence factor (see Ref.~\cite{Saad2003})
\[
  \rho \coloneqq \frac{\sqrt{\lambda_\text{max}/\lambda_\text{min}} - 1}{\sqrt{\lambda_\text{max}/\lambda_\text{min}} + 1} = 0.796,
\]
and thus rapid convergence of an iterative Krylov subspace method, like the Flexible Generalized Minimal Residual Method \cite{Saad1993} (known as FGMRES). See \cite{Praetorius2014} for a similar analysis on a preconditioner for the discretized PFC equation. 

Note, if the operator $\mathcal{G}^j$ contains higher order derivatives, a similar procedure for the construction of the preconditioner can be performed that may lead to a slightly different sequence of mass and diffusion-like systems but follows the same structure. A construction of a preconditioner for a sixth order PFC, an eighth order PFC, and a Lifschitz-Petrich type energy is discussed in \cite{PraetoriusThesis2015}, eventually leading to a sequence of diffusion equations to solve in the application of the preconditioner. The state potential $f^s$ may be extended in different directions and we expect the same procedure to work as above, provided that $f^s$ contains no derivative terms and the problem is well defined.

% -----------------------------------------------------------------------------------------------------------
\subsection{Mesh adaptivity}
The variables to solve for within the APFC model, namely $\eta_j$'s, are constant for relaxed crystals, oscillate with different periodicity according to the local distortion of the crystal with respect to the reference one, and exhibit significant variation at defects and solid-liquid interfaces. This allows for exploiting mesh adaptivity in order to optimize the numerical approach \cite{AthreyaPRE2007,Bercic2018}.
We set a local grid refinement in order to resolve the oscillation of amplitudes within grains and ensuring a proper resolution at defects and interfaces. 

\begin{figure*} % Fig. 2
\center
\includegraphics[width=\linewidth]{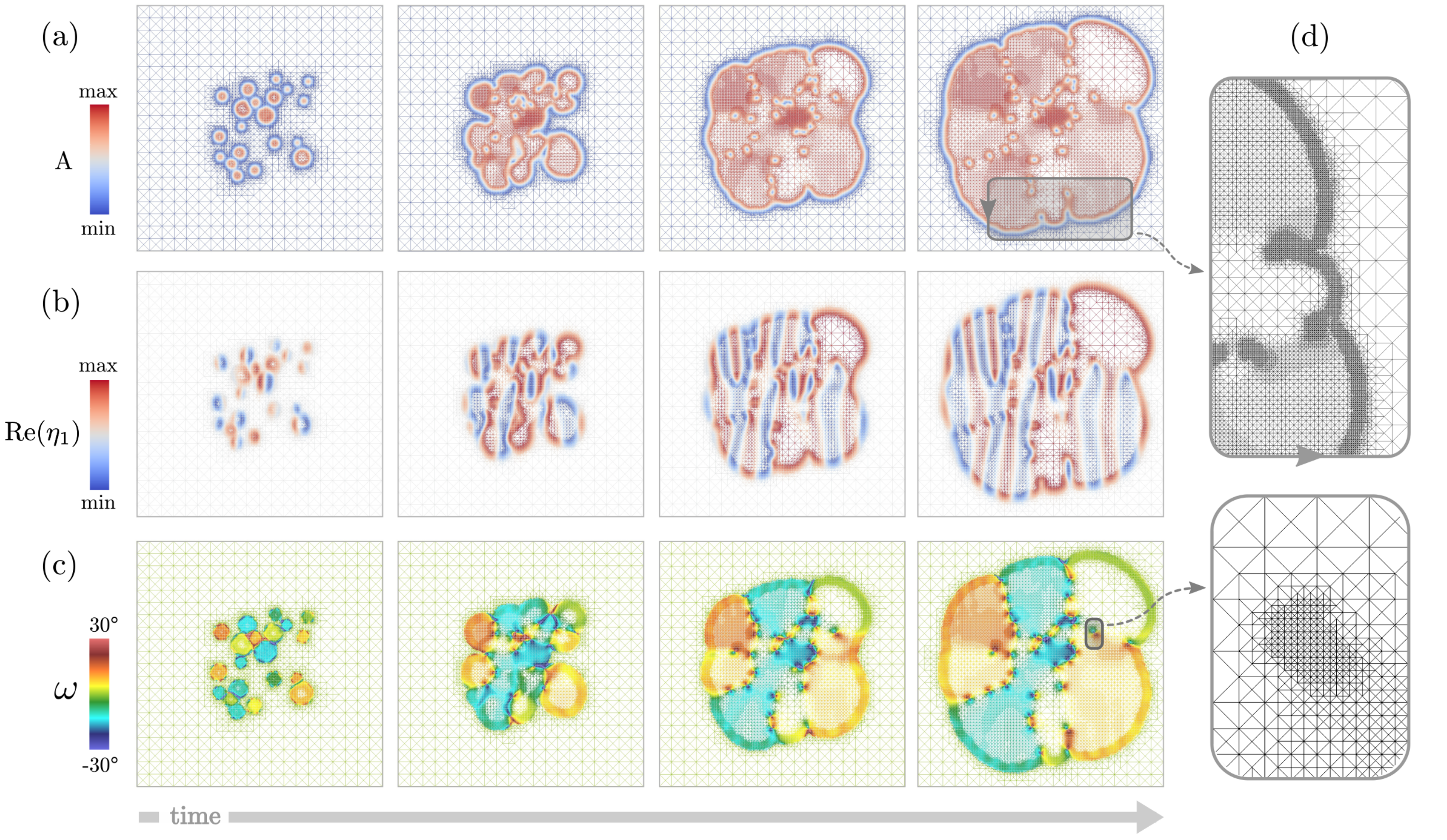} 
\caption{\label{fig:figure2}(Color online) Growth of 20 crystal seeds (in 2D) with $\theta \in (-15^\circ , 15^\circ)$ randomly distributed and triangular lattice symmetry. The spatial discretization is represented by means of the mesh while colors represent: (a) $A$, (b) $\text{Re}(\eta_1)$, (c) $\omega$. (d) Magnification of two regions showing the mesh on a smaller length scale at the solid-liquid interface (top) and at a defect (bottom). In this simulation the discretization bounds are $h_{\rm int} \approx 2.0$ and $h_{\rm max}\approx 40.0$.}
\end{figure*}

Amplitude functions describing a rotated crystal of an angle $\theta$ can be computed as
\begin{equation}\label{eq:amprot}
  \eta_j=\phi_j e^{ i\delta \mathbf{k}_j(\theta) \cdot \mathbf{r} },
\end{equation}
with $\delta\mathbf{k}_j(\theta)=\mathbf{k}_j \cdot \mathbf{R}(\theta) - \mathbf{k}_j$, $\mathbf{R}(\theta)$ the counterclockwise rotation matrix and $\phi_j$ the real amplitudes describing a relaxed, unrotated crystal. Therefore, the wavelength of $\eta_j$'s in the presence of a certain rotation $\theta$ is $\lambda_j(\theta)=2 \pi / |\delta\mathbf{k}_j(\theta)|$ and the spatial discretization is set as a fraction of the smallest $\lambda_j(\theta)$ i.e. h=$\min_j{(|\lambda_j(\theta)|)}/n$. We determined that a proper discretization is achieved with $n\geq 10$. In order to practically use this idea we then need to evaluate the local rotation field during the evolution. Notice that Eq.~\eqref{eq:amprot} cannot be just inverted due to its functional form. In order to compute the local rotation we then use the approach described in \cite{Salvalaglio2018a} where the rotational tensor $\boldsymbol{\omega}$ is determined by considering the curl of the local displacement with respect to the relaxed crystal. In practice, its components $\omega_{ij}$ representing the rotation in the $x_i$-$x_j$-plane, are given by
\begin{equation}
  \omega_{ij}=\frac{1}{2}\left[\frac{\partial u_i}{\partial x_j}-\frac{\partial u_j}{\partial x_i}\right],
  \label{eq:rot}
\end{equation}
with $u_i$ the result of inverting the system of equations 
\begin{equation}
  \mathbf{k}_j \cdot \mathbf{u} = \arctan\left[ \frac{\text{Im}(\eta_j)}{\text{Re}(\eta_j) }\right].
\end{equation} 
Explicit expressions for $u_i$, both in 2D and 3D can be found in Ref.~\cite{Salvalaglio2018a}. In 2D, Eq.~\eqref{eq:rot} delivers just a scalar field $\omega_{ij}\equiv \omega$. In 3D, $\boldsymbol{\omega}$ has three independent components, $\omega_d $ with $d=1,2,3$, and the largest rotation component is chosen in order to ensure the proper resolution for the fastest oscillation in the system. Within grains the discretization, $h_{\rm amp}$, then reads
\begin{equation}
  h_{\rm amp}=\frac{1}{n}\min_j{\left( \left| \lambda_j [ \max_d (\omega_d) ] \right| \right) }.
\end{equation}
The evaluation of rotations considered here is well posed in the solid phase and also at defects \cite{Salvalaglio2018a}. However, it is not well posed for the liquid, disordered phase as $\eta_j$'s vanish. Therefore, in order to properly discretize interface regions we define an additional refinement for the interfaces controlled by $h_{\rm int}$ where $|\nabla A|$ is significantly larger than a relatively small threshold $\epsilon$ as done in \cite{SalvalaglioAPFC2017}. As a result, $h_{\rm int}$ is ensured also at defects and we set it as the smallest resolution imposed in the system. In addition, a large discretization bound $h_{\rm max}$ is defined for region where $A \sim 0$ or where the local rotation vanishes. Summarizing these ideas, the local discretization, $h$, is set as
\begin{equation}\label{eq:hhh}
  h=\begin{cases}
  h_{\rm int}, & \mbox{if } |\nabla A| \geq \epsilon  \\
  \min(\hspace{1pt}\max(h_{\rm amp},h_{\rm int})\hspace{1pt},\hspace{1pt} h_{\rm max} \hspace{1pt}), & \mbox{if } A > 0 \mbox{ and } |\nabla A| < \epsilon \\
  h_{\rm max},& \mbox{elsewhere.} 
  \end{cases}
\end{equation}
Although not addressed here, this criterion can be extended in order to account for amplitude oscillations due to strain fields, exploiting continuous strain-field components, as derived in Ref.~\cite{Salvalaglio2018a}, instead of $\boldsymbol{\omega}$. 

The model and the mesh discretization are illustrated in \autoref{fig:figure2} where a simulation reproducing the growth of 20 crystal seeds in 2D with $\theta \in (-15^\circ , 15^\circ)$ is illustrated. A square simulation domain with a side length of $200\pi$ is considered (more details are reported in the following section). The initial rotation of the grain is set by initializing the amplitudes by means of \autoref{eq:amprot}. $A$, $\text{Re}(\eta_1)$ and $\omega$ at different stages during the evolution over time are shown superposed to the mesh used for the simulation. The refinement of the mesh where $A$ changes can be recognized at the solid-liquid interface and at defects, see in particular \autoref{fig:figure2}(a). The variables to solve for are the amplitudes $\eta_j$ as illustrated in \autoref{fig:figure2}(b) by $\text{Re}(\eta_1)$. Notice that the correct resolution of these oscillating functions is ensured by the discretization \eqref{eq:hhh}, exploiting the local rotations shown in \autoref{fig:figure2}(c).

The mesh adaptivity criterion illustrated in \autoref{fig:figure2} leads to larger refined regions and, in principle, to larger computational costs than the one reported in Ref.~\cite{AthreyaPRE2007} where an efficient mesh refinement strategy exploiting a polar representation of amplitudes has been proposed. However, as discussed therein, a robust regularization scheme for phase equations and special care to treat high-order derivatives were needed. This practically restricts the applicability of that method and prevents its extension to 3D calculations. An improvement of this description has been recently proposed in Ref.~\cite{Bercic2018}, exploiting spatially-dependent rotation of the $\mathbf{k}_j$ vectors, although it has been demonstrated only in 2D. The approach we use here allows for an optimized spatial discretization by using the standard APFC model. As it will be illustrated in the following it can be readily exploited to simulate different systems regardless of system dimensionality.

% -----------------------------------------------------------------------------------------------------------
\section{Performance studies}
For the numerical evaluation of the linear solvers and the mesh adaption strategy, we concentrate on two types of setups, a triangular symmetry in 2D (TRI) and a FCC symmetry in 3D. The domain $\Omega$ is of size $[0,s\cdot 10\pi]^d$ where $d$ is the space dimension and $s$ a scaling factor. The unit of lenght and time are dimensionless as they enter in the equations of the APFC model reported above. For the setups (TRI, FCC 1, FCC 2, FCC 3), we have chosen $s$ as ($20, 7, 14, 28$), respectively. To test the numerical framework we consider configurations already known and discussed in literature. The 2D triangular setup is initialized with 20 randomly positioned and oriented grains with orientation in the range $(-15^\circ, 15^\circ)$, corresponding to the configuration for the growth of a polycrystal as shown in \autoref{fig:figure2} (see, e.g., Refs.~\cite{AthreyaPRE2007,Spatschek2010,Salvalaglio2018a}). The 3D, FCC configurations, are initialized with one grain of size $30\pi$, $60\pi$, or $120\pi$, for the setup FCC 1, FCC 2, or FCC 3, respectively, rotated by $10^\circ$ about the [111] direction with respect to a surrounding, unrotated crystal (see Ref.~\cite{Salvalaglio2018}). The discretization bounds are $h_{\rm int} \approx 2.0$ and $h_{\rm max} \approx 40.0$ in 2D and $h_{\rm int} \approx 3.0$ and $h_{\rm max}\approx 60.0$ in 3D.

\begin{figure}[ht] % Fig. 3
\begin{center}
\includegraphics[scale=1]{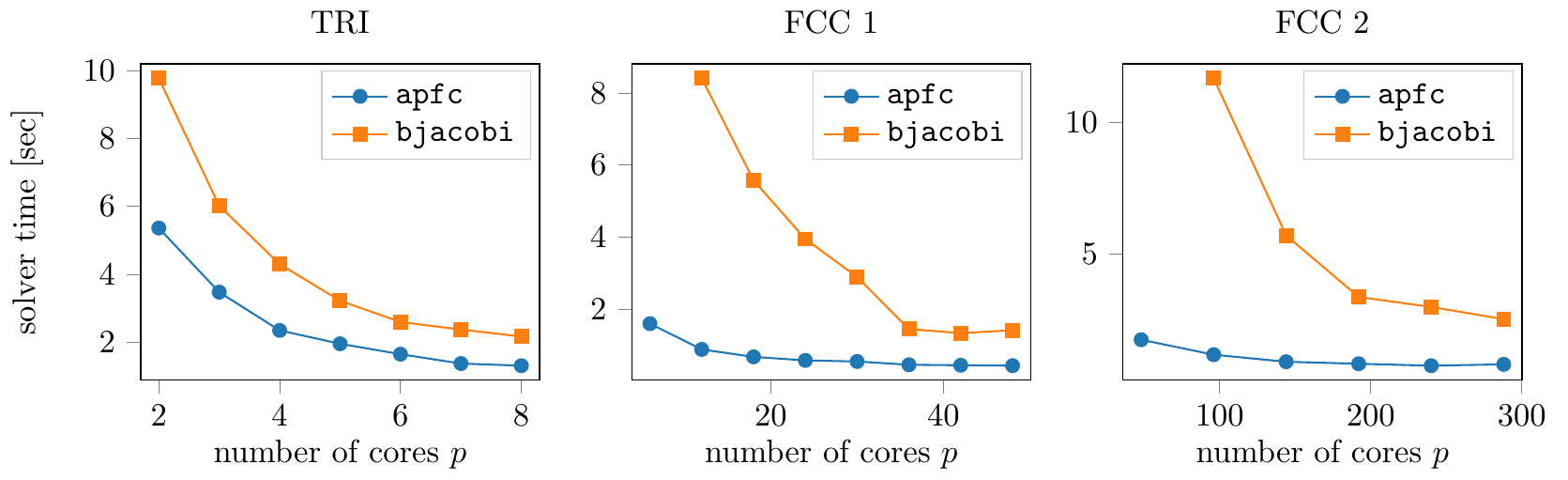}
\caption{\label{fig:figure3}(Color online) Comparison of \bjacobi\ and \apfc\ preconditioner for various setups. The large setup FCC 3 could not be solved using \bjacobi, due to memory limitations and is thus not shown here. The TRI setup is solved with a timestep size $\tau=2$ and the FCC setups with a timestep size $\tau=1$.}
\end{center}
\end{figure}

In a first comparison, we evaluate the linear-solver performance for the preconditioner $\mathbf{P}$ defined in Eq.~\eqref{eq:preconditioner} (\apfc) and a block Jacobi preconditioner (\bjacobi) applied to the blocks resulting from an element-wise domain decomposition, with a local sparse direct solver UMFPACK \cite{Davis2004}. The domain decomposition is combined with a distributed memory parallelization. In \autoref{fig:figure3} the timings for solving the linear system per timestep are summarized for TRI, FCC 1, and FCC 2 setup. Timings for the two preconditioners are evaluated by averaging over 100 timesteps over all amplitudes. Depending on the size of the setup different improvements of the \apfc\ preconditioner compared to the \bjacobi\ preconditioner can be found, ranging from a factor $10$ for small number of cores and thus large local partitions, to $2$ for large number of cores and thus small local partitions. In 2D the improvement is nearly constant over all subdomain sizes, but in 3D the local linear system size matters for the \bjacobi\ solver significantly. The three setups show a similar behaviour: the time to solve one linear system goes down with increasing number of processors, while it stagnates after some threshold in the number of cores. This is due to the fact that the local domain sizes become too small and additionally the load balancing gets more and more complicated. Increasing the number of cores further would increase also the solver time. Then, communication cost imbalance of the local problems would dominate the solution procedure.

\begin{figure}[ht] % Fig. 4
\begin{center}
\includegraphics[scale=1]{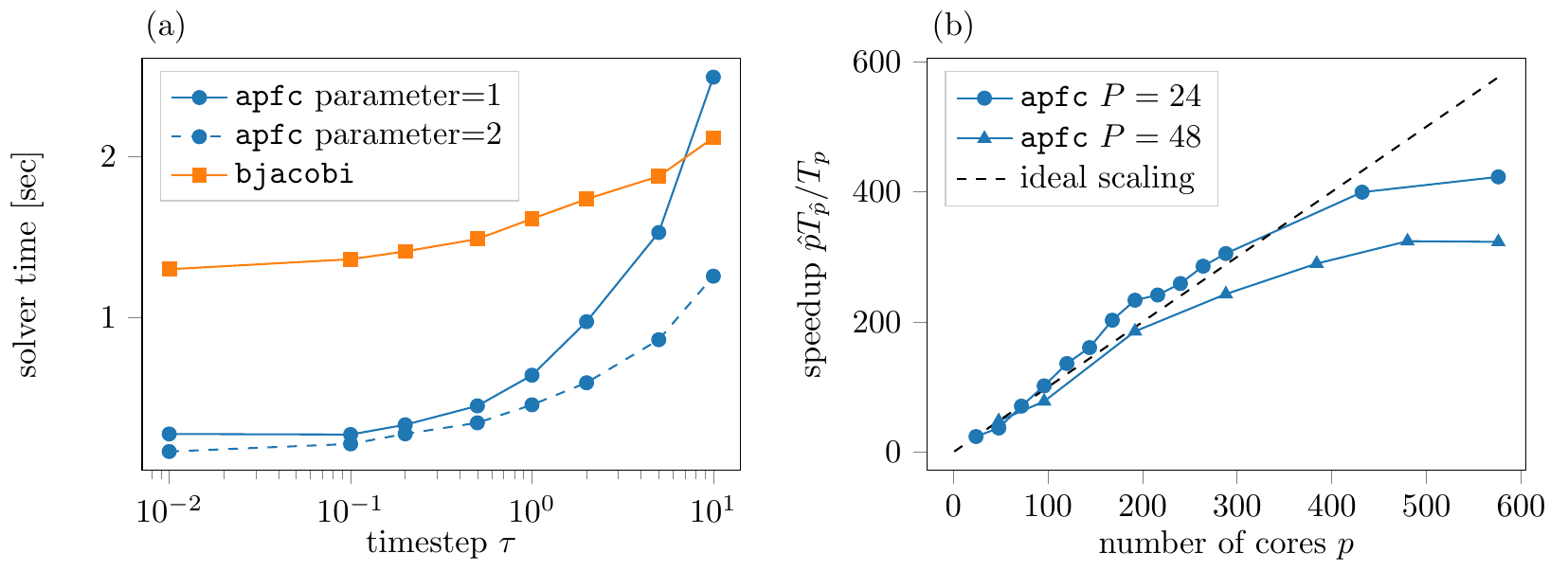} 
\caption{\label{fig:figure4}(Color online) \textbf{(a)} Comparison of the solver performance per timestep of setup FCC 1 for different timestep widths $\tau$. While the block preconditioner (\apfc) performs better if the timestep is small, the \bjacobi\ preconditioner seems more robust for large timesteps. The configuration {parameter=1} indicates a CG solver for the diffusion sub-system and {parameter=2} an AMG subsolver.
\textbf{(b)} Strong parallel scaling for the FCC 3 setup. Computations are run on compute nodes with $2\times 24$ cores each. Those nodes are either fully assigned ($P=48$) or with only half the cores used ($P=24$), resulting in different memory throughput and thus different performance. The minimal configuration $p_1$ for the speedup estimation is one node with 48 or 24 cores, respectively. The overstepping of the ideal solution line might be due to variations in the domain decomposition and thus in the communication pattern for different core counts. For large number of cores, the efficiency drops down to 60\%. This may be because of too small local problems.}
\end{center}
\end{figure}

While the \bjacobi\ preconditioner uses a direct solver on each local subdomain, that results in large memory costs, the \apfc\ preconditioner requires the solution of mass matrix and diffusion like equations, using an iterative solver optimized for the specific type of linear system. For the mass matrix, we use three iterations of a diagonally preconditioned conjugate gradient (CG) method. The diffusion system, on the other hand, is solved using either a diagonally preconditioned CG method with 5 iterations (parameter=1), or an algebraic multigrid method (AMG, parameter=2), based on the so-called Hypre BoomerAMG \cite{Henson2002}, with one V-cycle and symmetric relaxation. In \autoref{fig:figure4}(a) the difference in the solver time per timestep is plotted, depending on the timestep parameter $\tau$ that is part of the preconditioner definition \eqref{eq:preconditioner}. Increasing the timestep size results in slower convergence of all the linear solvers, but the effect is more pronounced with the \apfc\ preconditioner. In the same figure also the difference in the solver time for different solver parameters is shown. For this setup, a simple CG iteration outperforms the AMG subsolver. Concerning just the performance of the linear solver the optimal timestep would be as large as possible, but the accuracy of the scheme is just first order in $\tau$ so that a balance between accuracy and performance has to be found for a concrete setup.

A third benchmark considers the parallel scalability of the \apfc\ preconditioner. Therefore, we run setup FCC 3 with an increasing number of subdomains, that is, an increasing number of compute cores. These benchmarks are run on the JUWELS cluster of the J\"{u}lich Supercomputing Centre on regular $2\times 24$ core nodes. Since an allocation of a different number of running processes per node results in different node performance, e.g., due to parallel usage of memory pipelines, we have assigned a fixed number of cores per node for each scaling benchmark. In \autoref{fig:figure4}(b) the relative speedup for an allocation of 24 and 48 cores per node is plotted with respect to the smallest possible number of cores, i.e., one node with 24 or 48 cores, respectively. Using less cores per node leads to a slightly better speedup that might even exceed the ideal scaling line. Although the parallel efficiency drops down if the local problems get too small, in the intermediate range of up to 300 cores, we see very good speedup and thus parallel efficiency of the \apfc\ preconditioner.

% -----------------------------------------------------------------------------------------------------------
\section{Application: Growth of polycrystals with dislocations}
In this section we illustrate some applications of the numerical approach illustrated above. In order to allow for estimates and comparisons in terms of number of atoms, e.g. to other methods, simulations or real systems, the length scale is here reported in terms of the lattice constant for the considered symmetries as set by the corresponding set of $\mathbf{k}_j$ vectors. They read $a_{\rm tri}$ and $a_{\rm fcc}$ for the triangular and FCC lattice symmetry, respectively.  
\begin{figure*} % Fig. 5
\center
\includegraphics[width=\linewidth]{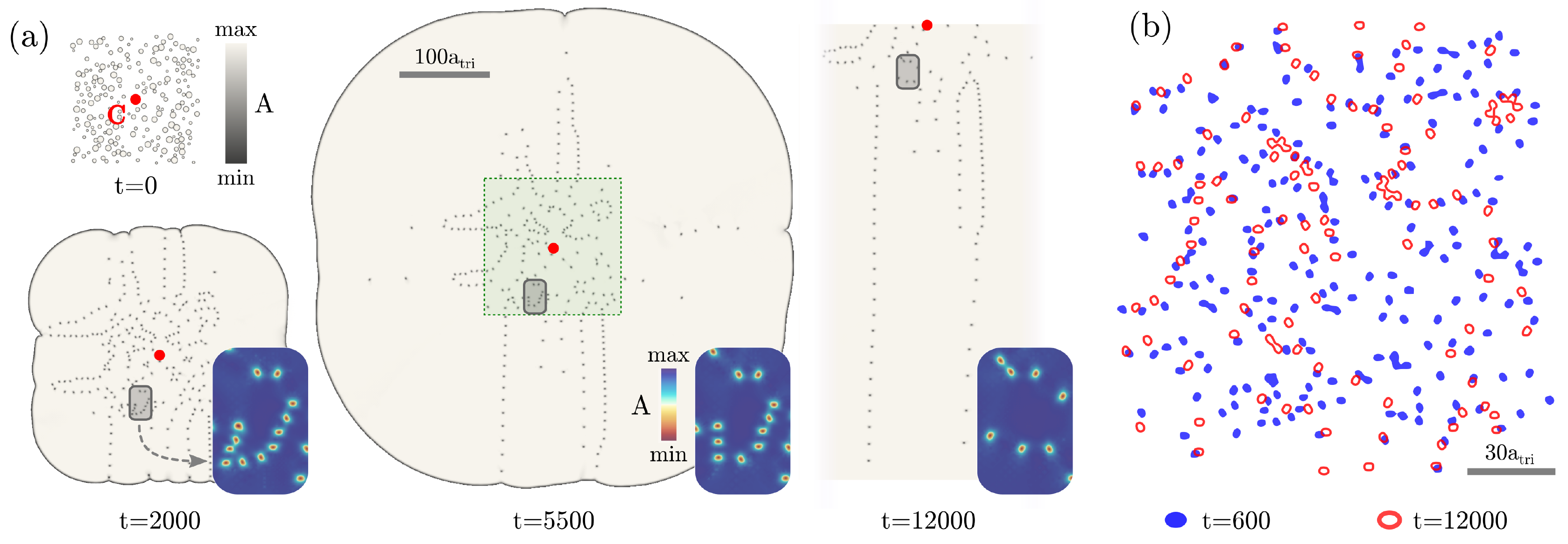} 
\caption{\label{fig:figure5}(Color online) Large-scale simulation of the growth of a polycrystal in 2D with triangular symmetry. The initial configuration consists of 200 crystal seeds with rotation $\theta \in (-15^\circ , 15^\circ)$, randomly distributed around center of the simulation domain (marked by the red point C). (a) The initial configuration ($t=0$) and three representative steps are shown in terms of the region where $A>0$, i.e. in term of the solid phase. $t=12000$, where the crystal fills the entire simulation domain, is reported by means of a portion of the crystal highlighting the formation of straight GBs. $A$ is also shown by greyscale map, showing the presence of defects. The length scale is the same for every step. Insets show the arrangement of defects in a small portion of the crystal (gray shaded area, see t=2000). (b) Comparison between the distribution of defects in the green shaded region of panel (a) at $t=600$ (blue, filled) and $t=12000$ (red, empty).}
\end{figure*}
\autoref{fig:figure5} shows the growth of 200 crystal seeds in 2D having triangular symmetry. A rotation of crystals $\theta \in (-15^\circ , 15^\circ)$ set as initial condition by means of \autoref{eq:amprot} as in \autoref{fig:figure2} is considered. We set here a square domain of side length $\sim 10^3 a_{\rm tri}$. The initial crystal seeds are generated in a square region at the center (see \autoref{fig:figure5}(a), $t=0$). As shown by two representative steps during the evolution, $t=2000$ and $t=5500$, the growth of these seeds results in a polycrystal with several dislocations at the center while almost straight GBs form between the peripheral grains which are free to grow towards the liquid phase. The formation of such GBs is also highlighted at $t=12000$ by means of a portion of the entire simulation domain. Insets of \autoref{fig:figure5}(a) illustrate the arrangement of defects on a smaller length scale in a portion of the crystal. The motion and eventual annihilation of dislocations is accounted for by the approach as illustrated in \autoref{fig:figure5}(b) where the arrangement of defects in the central region of the simulation (green shaded square superposed to the crystal at $t=5500$) is compared at $t=600$ (blue, filled) and $t=12000$ (red, empty). The coarsening dynamics results faster in the early stages than at later times as can be noticed by comparing the main features of the dislocation networks at different times in \autoref{fig:figure5}. Indeed, small grains are present at the beginning leading to the formation of curved grain boundaries formed by a few dislocations, which, in turn, move fast \cite{Mullins1956,Doherty1997}. Later, the grains at the center of the polycrystal are larger and the resulting GBs are more stable having smaller curvatures. However, the coarsening dynamics with motion and annihilation of defects continues as can be notice in the insets of  \autoref{fig:figure5}(a), where significant changes in the arrangement of defects are observed at later times. The straight grain boundaries forming due to the growth of the polycrystal can be considered as long-lasting defects as their curvature is negligible, while the spacing between dislocations depends on the relative tilts of grains. Notice that the growth velocity of the polycrystal is a function of the parameters entering the free energy that control the energy difference between the solid and the liquid phase \cite{Elder2007}. Therefore, faster or slower grain growth compared to defect motion can be inspected.

The main features of the dynamics obtained by the standard APFC approach considered here qualitatively correspond to predictions of classical theories and simulations \cite{Mullins1956,Doherty1997}. However, a quantitative description can be achieved by accounting for extensions of the PFC and APFC approach, including a proper description of elastic and plastic relaxation during the motion of defects \cite{Heinonen2016,Skaugen2018b}.

\begin{figure*} % Fig. 6
\center
\includegraphics[width=\linewidth]{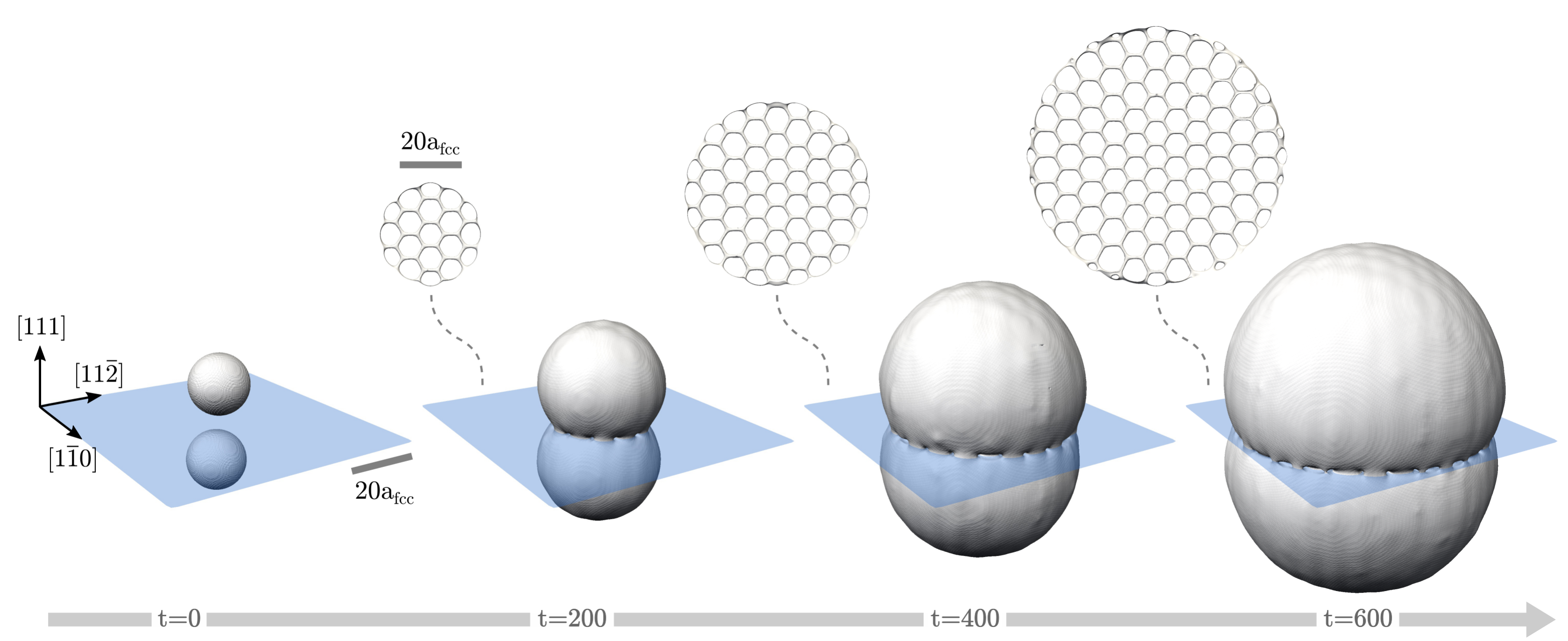} 
\caption{\label{fig:figure6}(Color online) Growth of 2 crystal seeds in 3D with FCC lattice symmetry and rotation $\theta =\pm 5^\circ$ about the [111] directions. Four steps during the evolution are shown by means of region where $A>0$. Insets illustrate the the defect structure forming at the planar, twist (111) GB by means of the regions at the blue shaded plane where $0<A<0.8\max_{\Omega}(A)$.}
\end{figure*}

\begin{figure*} % Fig. 7
\center
\includegraphics[width=\linewidth]{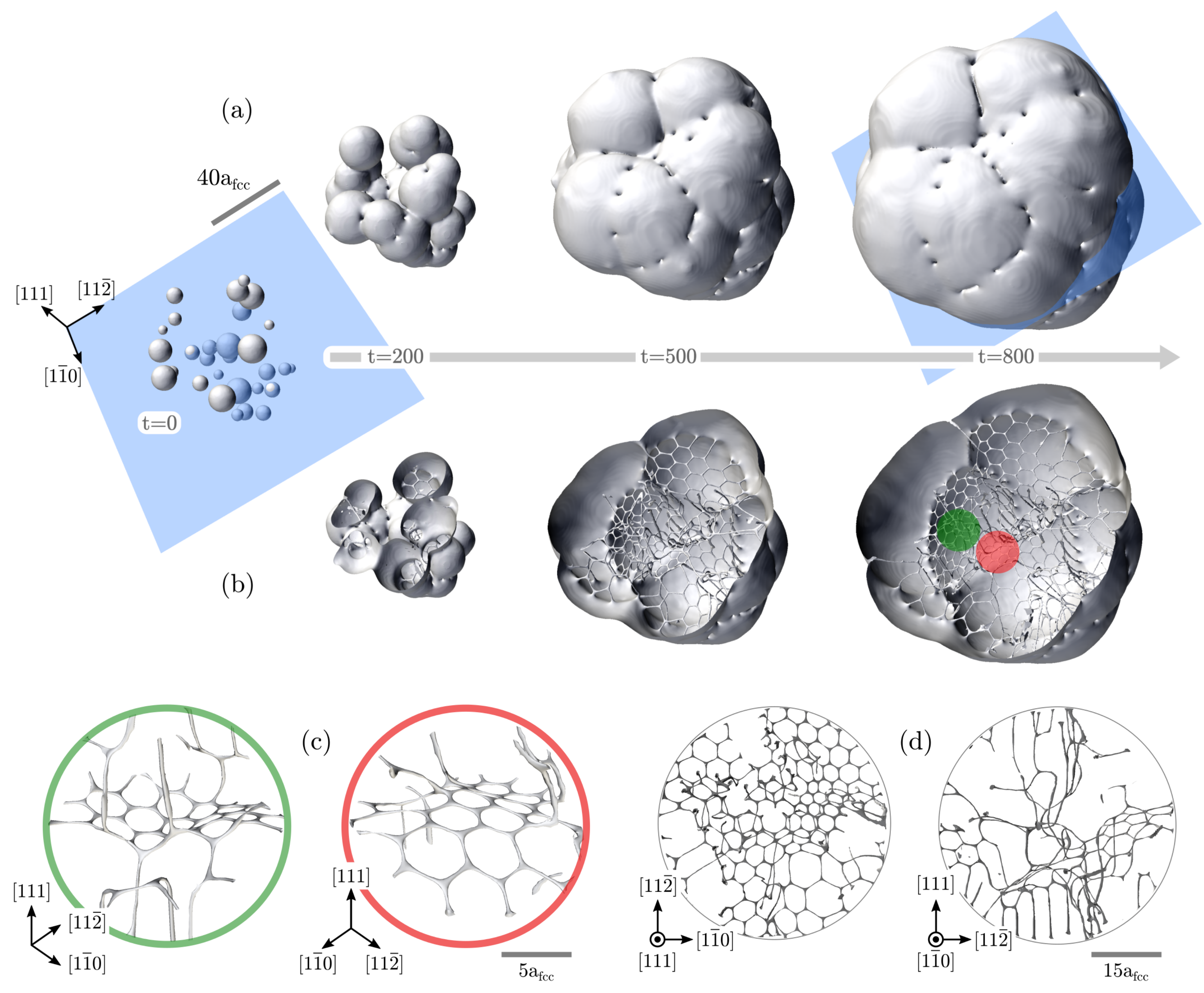} 
\caption{\label{fig:figure7}(Color online) Growth of 30 crystal seeds in 3D with FCC lattice symmetry and random rotation $\theta \in (-15^\circ , 15^\circ)$ about the [111] directions. Initial seeds are randomly distributed at the center of the simulation domain. (a) Regions where $A>0$. (b) Isosurface $A=0.8\max_{\Omega}(A)$ in half of the domain, cut along the (111) plane passing through its center (blue shaded plane illustrated at $t=0$ and $t=800$). (c) Magnification of two small spherical regions inside the polycrystal showing defect arrangements on a small length scale (see corresponding colors in panel (b), t=800). (d) View of hemispheres, fully contained in the growing polycrystal, showing the arrangement of defects from two different perspectives: perpendicular to the [111] direction (left) and to the [1$\bar{1}$0] direction (right).}
\end{figure*}

A 3D example is shown in \autoref{fig:figure6}. Two FCC crystals, having a rotation of $\pm 5^\circ$ about the [111] direction are considered. The simulation size is set as in the setup FCC 3, corresponding to a square domain of side length $\sim 80 a_{\rm fcc}$. They are also aligned along the [111] direction, therefore their growth is expected to form a (111) twisted GB with a typical hexagonal arrangement of defects \cite{Scott1981,DeHosson1990,Salvalaglio2018}. \autoref{fig:figure6} reports four stages during the evolution in terms of $A>0$ along with the dislocation networks at the resulting planar GB illustrated as insets. They correspond to regions where $0<A<0.8\max_{\Omega}(A)$ at the (111) blue shaded plane. The study of these kind of planar GBs, in particular for what concern the morphology of the emerging dislocation network, are typically accounted for by assuming ideally infinite crystals by means of periodic boundary conditions or even just by 2D approaches. Here, the GB is obtained together with the explicit description of crystal growth which enable more general investigations tackling the simultaneous presence of different GBs with different orientations. A more general case, illustrating the general capability of the approach, is reported in \autoref{fig:figure7}. Therein 30 crystals having random rotation about the [111] direction are considered. A simulation domain that is double the size of the setup FCC 3, namely corresponding to a square domain of side length $\sim 160 a_{\rm fcc}$, is considered here. \autoref{fig:figure7}(a) shows the morphologies of the seeds and of the resulting, growing polycrystal in terms of $A>0$ regions. \autoref{fig:figure7}(b) shows half of the crystals reported in \autoref{fig:figure7}(a) by the isosurface $A=0.8\max_{\Omega}(A)$ revealing also the dislocation network forming at the internal GBs. In this case the initial seeds are distributed randomly and are not aligned along a specific direction. Therefore, together with the twist GBs as shown in \autoref{fig:figure6},  other orientations for the boundaries between grains are present thus leading to different morphologies for the dislocation networks. This can be observed in more detail in \autoref{fig:figure7}(b)--(d). Hexagonal patterns mostly lying on (111) planes, can be recognized along with elongated defects typical of pure tilt GBs. Moreover, similar patterns having different spacing between dislocations are present due to different relative rotations between grains.

The simulations reported in this section are unprecedented in terms of sizes for what concern APFC (and PFC) approaches. Just by focusing on the largest system of Fig.~\ref{fig:figure7}, we were able to simulate a crystalline system with FCC lattice symmetry including $\sim 8 \cdot 10^6$ atoms. For specific materials exhibiting such a lattice symmetry as, e.g, Cu, Ag/Au, Pb this would mean a volume of $\sim (45\ {\rm nm})^3$, $\sim (50\ {\rm nm})^3$, $\sim (65\ {\rm nm})^3$ respectively \cite{Davey1925}, which lie in the typical size range of nanoparticles and nanostructures. The simulation in Fig. 7 is done on 720 cores. Increasing the number of cores will also allow to consider even larger samples and thus realize the envisioned multiscale approach, ranging from atomistic details to micrometer sizes.

% -----------------------------------------------------------------------------------------------------------
\section{Conclusions}
We illustrated a numerical approach to solve the equations of the APFC model efficiently. A specific discretization scheme combined with a nonlinear solver has been proved to allow for unprecedented size and performances for both 2D and 3D APFC simulations. In particular, we have constructed a schur-complement preconditioner for iterative Krylov-subspace methods that outperforms a block Jacobi solver for the linear systems arising from the discretization with adaptive finite elements. The schur-complement solver requires much less memory, converges fast in terms of wall-clock time, and scales well in parallel setups. On the other hand, it is more sensitive to an increase in timestep size, compared to the \bjacobi\ solver. 

Moreover, an optimized criterion for mesh adaptivity has been proposed and used, exploiting features of the complex amplitude functions $\eta_j$ and derived physical quantities such as the local rotation field. This can be applied in both 2D and 3D and allows for a significant reduction of the overall number of DOFs.

Some applications involving the growth of polycrystals in 2D and 3D as well as the simultaneous description of dislocations forming at GBs have been shown. They set new limits for APFC and then, more in general, PFC approaches, enabling the investigation of large systems matching the size of real nanostructures. This is a crucial step to address large, mesoscale problems still retaining details of the atomic length scale. Future work will be devoted to explicitly include further details, compatible with the APFC model, in the numerical framework presented here as, for instance, an improved description of interface-energy anisotropy \cite{Ofori-Opoku2018,Salvalaglio2015}, binary systems \cite{Elder2010a}, an improved description of the dynamics \cite{Heinonen2016,Skaugen2018b}, the coupling with magnetic fields \cite{Backofen2018}, as well as improvements on scaling properties of the numerical approach to enable larger systems addressing grain growth in 3D.

% -----------------------------------------------------------------------------------------------------------
\section{Acknowledgements}
The authors gratefully acknowledge the Gauss Centre for Supercomputing e.V. (www.gauss-centre.eu) for funding this work by providing computing time through the John von Neumann Institute for Computing (NIC) on the GCS Supercomputer JUWELS at J\"ulich Supercomputing Centre (JSC), under the grant No. HDR06. A.V. acknowledge the financial support from the German Research Foundation (DFG) under Grant No. SPP 1959.

\end{document}